\documentclass[conference]{IEEEtran}
\IEEEoverridecommandlockouts
\usepackage{cite}
\usepackage{amsmath,amssymb,amsfonts}
\usepackage{algorithmic}
\usepackage{graphicx}
\usepackage{textcomp}
\usepackage{epsfig}
\usepackage{epstopdf}
\usepackage{xcolor}
\def\BibTeX{{\rm B\kern-.05em{\sc i\kern-.025em b}\kern-.08em
    T\kern-.1667em\lower.7ex\hbox{E}\kern-.125emX}}

\def\be{\begin{equation}}
\def\ee{\end{equation}}

\def\mih{\mathbf{h}}

\def\mix{\mathbf{x}}

\def\mi0{\mathbf{0}}

\begin{document}

\title{Artificial Neural Network for Resource Allocation in Laser-based Optical wireless Networks 
{\footnotesize \textsuperscript{}}
 \thanks{This work  has been supported by the Engineering and Physical Sciences Research Council (EPSRC), in part by the INTERNET project under Grant EP/H040536/1, and in part by the STAR project under Grant EP/K016873/1 and in part by the TOWS project under Grant EP/S016570/1. All data are provided in full in the results section of this paper.}

}
\author{\IEEEauthorblockN{Ahmad Adnan Qidan, Taisir El-Gorashi1, Jaafar M. H. Elmirghani}
\IEEEauthorblockA{School of Electronic and Electrical Engineering, University of Leeds, LS2 9JT, United Kingdom 
\\Email: \{a.a.qidan, t.e.h.elgorashi, j.m.h.elmirghani\}@leeds.ac.uk}  


}

\maketitle

\maketitle

\begin{abstract}
Optical wireless communication offers unprecedented communication speeds that can support the massive use of the Internet on a daily basis. In indoor environments, optical wireless networks  are usually  multi-user multiple-input multiple-output (MU-MIMO) systems, where a high number of optical access points (APs) is required to ensure coverage. In this work, a laser-based optical wireless network  is considered for serving multiple users. Moreover, blind inference alignment (BIA) is implemented to achieve a high degree of freedom (DoF) without the need for  channel state information (CSI) at transmitters, which is difficult to provide in such wireless networks. Then, an objective function is defined  to allocate the resources of the network taking into consideration the requirements of users  and the available resources. This optimization problem can be solved through exhaustive search or distributed algorithms. However, a practical algorithm that provides immediate  solutions in real time scenarios is  required. In this context,  an artificial neural  network (ANN) model is derived in order to obtain a sub-optimal solution with low computational time. The implementation of the ANN model involves three important steps, dataset generation, offline training, and real time application. The results show that the trained ANN model  provides a significant solution close to the optimal one.  
\end{abstract}

\section{Introduction}
In recent years, researches have investigated exploiting  the enormous unlicensed-bandwidth of the optical band to support the increase in  user demands. In this context, optical wireless networks are investigated using light emitting diode (LED)  for providing illumination and data transmission. Despite the high achievable data rates in these networks compared to radio frequency (RF) wireless networks, the low modulation speed of LEDs limits the potential of considering  optical wireless notworks in the next generation (6G) of wireless networks. Infrared (IR) Lasers, namely Vertical-Cavity Surface Emitting (VCSEL) sources, are used as transmitters in \cite{1919,AA19901111}  under eye safety constraints. The features of VCSELs  including their  high modulation speed compared with LED, low power consumption and low cost  make them a strong candidate to achieve unprecedented rates that fulfill the escalating demands of users.     

In laser-based wireless networks, deploying a high number of optical access points (APs) is needed to ensure coverage. Therefore, these networks are naturally considered as multi-user multiple-input multiple-output (MU-MIMO) systems. In this sense,  transmit precoding schemes such as  zero forcing (ZF) \cite{QHQHALMH} are implemented to align multi-user interference with the need for channel state information (CSI) at transmitters. It is worth mentioning that the characteristics of the optical signal limit the performance of ZF where a DC bias current must be applied to ensure the non-negativity of the transmitted signal. In \cite{MMAL17,8636954} a reconfigurable optical detector is proposed with the aim of implementing blind interference alignment (BIA), which serves multiple users simultaneously without CSI at transmitters following a certain methodology. It is shown that the precoding matrices of BIA are given by positive values, and it is more suitable for optical wireless networks compared to ZF. 

Resource management  plays an important role in maximizing the overall sum rate of the network. In \cite{AA19901111}, an objective function is defined to allocate the resources based on the connectivity of users.
In \cite{9521837},  an optimization problem is formulated to maximize the sum rate of users by allocating their requirements of resources. In general, such optimization problems have  high complexity that increases considerably with the size of the network. Therefore, distributed algorithms using Lagrangian multipliers are used in order to reduce the complexity, while  providing sub-optimal solutions  \cite{AA19901111,9521837,FRLHANZO}. However, a distributed algorithm requires  an iterative algorithm, affecting its  accuracy in obtaining solutions with minimum errors in real time scenarios due to the fact that users might change their requirements, locations, etc., while  the algorithm performs resource allocation. Recently, artificial neural networks (ANNs)  have received massive interest as an alternative technique to solve the optimization problems in high complexity problems in low computational time \cite{sun20,luca19}, where an ANN model can be trained over a dataset in an offline phase, so that accurate  solutions can be obtained instantaneously in real time scenarios.    


In contrast to the work in \cite{9521837}, in this paper, an optimization problem is formulated in a laser-based wireless network to satisfy the requirements of users and maximize the utility function of the sum rate using a certain ANN model. We first derive the achievable  user rate considering BIA as a transmission scheme. Then, the optimization problem  is formulated under several constraints of user-requirements and AP-capacity limitations. This problem can be solved using exhaustive search to  provide an optimal solution with high complexity. Finally, an ANN model is defined with the aim of providing a sub-optimal solution with low complexity. The ANN model is implemented in three steps: Dataset generation, offline training, and real time scenario. The results show that the proposed ANN model provides a  solution  close to the optimal one. Moreover, BIA is more suitable for optical wireless network than ZF.    

\begin{figure}[t]
\begin{center}\hspace*{0cm}
\includegraphics[width=0.85\linewidth]{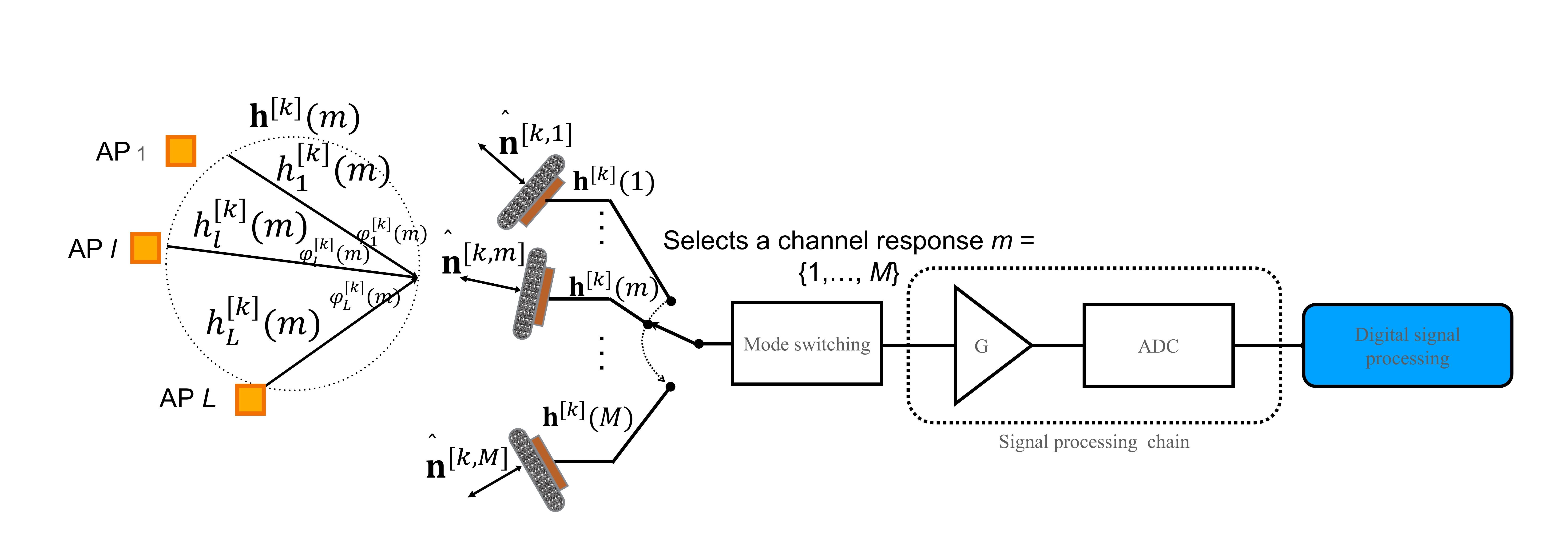}
\end{center}
\vspace{-2mm}
\caption{ Reconfigurable optical detector with $ M $ photodiodes.}\label{de}
\vspace{-2mm}
\end{figure}    
\section{System Model} 
\label{sec:system}
We consider a laser-based wireless network composed of $ L $, $ l=\{1, \dots, L\} $, APs serving $ K $, $ k=\{1, \dots, K\} $, users distributed on the receiving plane.  Each user is equipped with a reconfigurable detector, which consists of $ M $ multiple photodiodes as shown in Fig. \ref{de}, providing a wide field of view (FoV) to ensure the connectivity to most of the available APs. It is worth mentioning that  this detector has the ability to provide linearly independent channel responses where each photodiode has a distinct direction $ \mathbf{\widehat{n}} $, more details are in \cite{MMAL17,8636954}. In this sense, the  received signal of a generic user $ k $ at photodiode $ m $, $ m \in M $, is given by
\begin{equation}
y^{[k,l]}[n]=\mih^{[k]}(m^{[k]}[n])^{T} \mix[n]+  z^{[k]}[n],
\end{equation}
where $ \mih^{[k]}(m^{[k]}[n])^{T} \in \mathbb{R}_+^{L\times 1} $, $ m^{[k]}[n] $  is a preset mode selected by photodiode $ m $ at time slot $ n $, $ \mix $ is the transmitted signal and $ z^{[k,l]} $ is real valued additive white Gaussian noise with zero mean and variance given by the sum of shot noise,  thermal noise and the intensity noise of VCSEL. In this work, CSI is avoided, and all APs  are connected to a central unit (CU), which controls the resources of the network. Moreover, users can send  their resources requirements through a WiFi link.
\subsection{Transmitter}
The optical channel  between user $ k $ and AP $ l $  at photodiode $ m $ can be expressed as 
\begin{equation}
h^{[k,l]}(m)=h_{ \mathrm{LoS}}^{[k,l]}(m) + h_{\mathrm{diff}}(f) e^{-j2\pi f \Delta T}, 
\end{equation}
where $h_{ \mathrm{LoS}}^{[k,l]}(m)$ denotes  Line-of-Sight (LoS) components of the direct link, $h_{\mathrm{\mathrm{diff}}}$ is the diffuse channel (Non-LoS) and $\Delta T$ is the delay between LoS and diffuse components. Each VCSEL illuminates a small and confined area, and therefore, the diffuse component can be neglected,  for the sake of simplicity, where most of the received power is due to LoS components \cite{AA19901111}. 

The transmitted power of VCSEL can be determined  based on the  beam waist $ W_{0} $, the wavelength $ \lambda $ and the distance between the ceiling and the receiving plane $ d $. Moreover, the beam profile of the VCSEL transmitter is Gaussian ignoring higher weak modes, and its intensity is defined as a function of the radial distance $ r $ from the center of the beam spot and the distance $ d $, i.e., $ I(r,d) $. In this context, considering  the transmitted power of VCSEL $ l $,  $ { P_{t,l}} $, and its beam radius, $ W_{d} $, at distance $ d $, the received power by user $ k $ at photodiode $ m $ located right below VCSEL $ l $ can be expressed as 
\begin{equation}
\begin{split}
&P_{m,l}=\\
&\int_{0}^{A_m /2 \pi} I(r,d) 2\pi r dr = P_{t,l}\left[1- \mathrm{exp}\left(- 2 \left(\frac{ A_{m}}{2 \pi W_{d}}\right)^{2}\right)\right],
\end{split}
\end{equation}
where $ A_m = \frac{A_{rec}}{M} $ is   the area of photodiode $ m $,  assuming the whole area of the reconfigurable detector is $ A_{rec} $.   
  
\section{Blind Multiple Access scheme }
 \label{sec:Blind}
In \cite{Gou11,8636954},  BIA is proposed for  interference management in RF and optical wireless networks, respectively,  without the need for CSI at transmitters, . Basically,  in BIA,  a transmission block referred to as supersymbol is generated, which consists of two blocks, Block 1 and Block 2. In this section, the construction of the supersymbol is presented  first for a toy  example, and then, the achievable user rate is derived for the general case.

\begin{figure}[t]
\begin{center}\hspace*{0cm}
\includegraphics[width=0.4\linewidth]{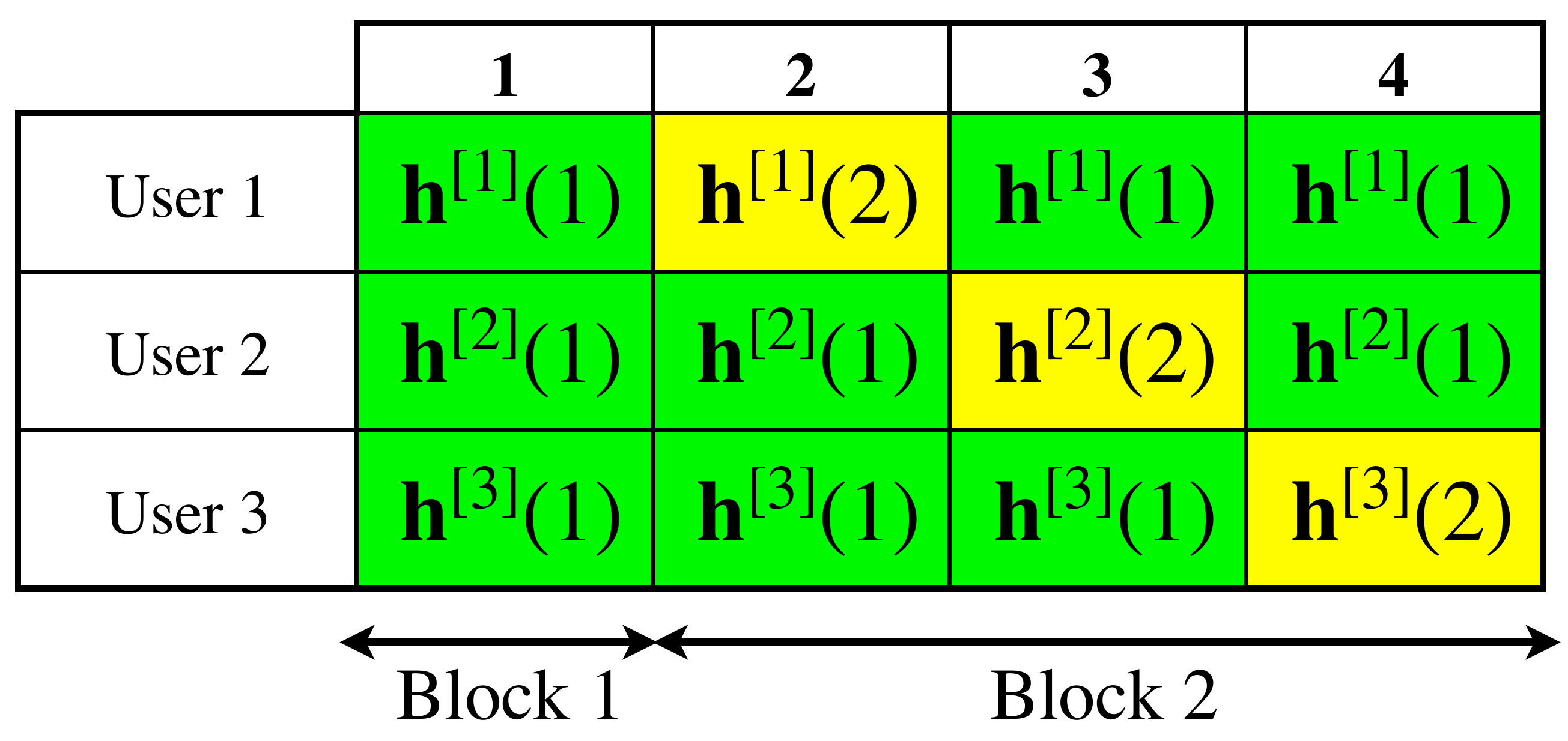}
\end{center}
\vspace{-2mm}
\caption{The supersymbol of BIA for $ L=2 $ and $ K=3 $. Each color represents a preset mode.}\label{bi}
\vspace{-2mm}
\end{figure}
Let us consider a case in which $ L=2 $  APs that serve $ K=3 $ users, each user is equipped with a reconfigurable photodetector. In BIA-based interference management, all users receive their  information over Block 1 of the supersymbol generating interference among them. However, orthogonal transmission is carried out over Block 2 providing enough dimensions for each user to measure and cancel the interference received over Block 1.  To guarantee this methodology for the case considered, Block 1 comprises one time slot, while three time slots belong to Block 2, forming the supersymbol of BIA as shown in Fig.\ref{bi}. In this context, the transmitted signal is given by
\begin{equation}
\label{T_si}
\mathbf{X} = 
\begin{bmatrix}
\mathbf{I} \\ \mathbf{I}  \\ \mathbf{0}  \\ \mathbf{0}  
\end{bmatrix}
\mathbf{u}^{[1]}
+
\begin{bmatrix}
\mathbf{I} \\ \mathbf{0}  \\ \mathbf{I}  \\ \mathbf{0}  
\end{bmatrix}
\mathbf{u}^{[2]}
+ 
\begin{bmatrix}
\mathbf{I} \\ \mathbf{0}  \\ \mathbf{0}  \\ \mathbf{I}  
\end{bmatrix}
\mathbf{u}^{[3]},
\end{equation}
where $\mathbf{u}^{[k]} = \begin{bmatrix}u_1^{[k]} & u_2^{[k]} \end{bmatrix}$ is the symbol transmitted to user $ k $. From equation \eqref{T_si}, orthogonal transmission among the users can be observed over the second, third and fourth time slots. Focusing on user 1, without loss of generality,
the interference because of the transmission to the other users over Block 1, i.e., the first time slot, is measured  over time slots 3 and 4 of Block 2, and then, it can be subtracted afterwords at the cost of increasing noise. As a consequence, user 1 decodes 2 DoF of $ \mathbf{u}^{[1]} $ transmitted over $ \{1,2\} $ time slots. Similarly, users 2 and 3 can decode  2 DoF of  $ \mathbf{u}^{[2]} $ and $ \mathbf{u}^{[3]} $, transmitted over $ \{1,3\} $ and $ \{1,4\} $ time slots, respectively. Therefore, the sum DoF  equals to $ 6/3 $ DoF for the case considered, which is higher than orthogonal transmission schemes such as TDMA.
 
For the general case where  $ L $  APs serve $ K $ users, the BIA supersymbol comprises $ (L-1)^{K}+K(L-1)^{K-1} $ time slots, more mathematical details are provided in \cite{Gou11,8636954}. In this sense, the achievable rate of user $k$ is
\begin{equation}
r^{[k]} = \frac{1}{L+K-1} \log_2\left(\mathbf{I} + P_{\rm{str}}\mathbf{H}^{[k]} {\mathbf{H}^{[k]}}^H {\mathbf{R}_z}^{-1} \right),
\end{equation}
where $P_{\rm{str}}$ is the power allocated to each stream, $\mathbf{H}^{[k]} = \begin{bmatrix} \mathbf{h}^{[k]}(1) & \dots & \mathbf{h}^{[k]}(L) \end{bmatrix}^T \in \mathbb{R}^{L\times L}$ is the channel matrix of user $k$, and $\mathbf{R}_z = \begin{bmatrix}K\mathbf{I}_{L-1} & 0 \\ 0 & 1 \end{bmatrix}$ is the covariance matrix of  noise.
\section{Problem Formulation}
Resource allocation schemes enhance  the performance of wireless networks in terms of resource utilization and the maximization of the overall sum rate. In particular, the resources of a network can be allocated uniformly among users avoiding complexity. However, this way might not satisfy the requirements of users. In this sense, a utility function-based optimization problem is formulated with the aim of maximizing the sum rate of the users by allocating their resource demands  taking into consideration the capacity limitations of the APs, as in the following  
\begin{equation}
\label{OP2}
\begin{aligned}
\max_{e} \quad &  \sum_{l\in L} \,\sum_{k\in K} \log \left(1+ \xi_{k} ~ e^{[k,l]} r^{[k,l]}\right)\\
\textrm{s.t.} \quad & \sum_{k\in K} e^{[k,l]} \leq \rho_l,\, \forall l \in L\\
\quad &  \sum_{l\in L}  e^{[k,l]} \leq e^{[k]}_{\max},\, \forall k \in K\\
 \quad & \sum_{l\in L}  e^{[k,l]}  \geq e^{[k]}_{\min}, \, \forall k \in K,
\end{aligned}
\end{equation}
where $U(\cdot)= \log(\cdot)$ \footnote{The objective function is considered in the form of  $  \log \left( 1+ \xi_{k} ~ e^{[k,l]} r^{[k,l]}\right)  $ in order to avoid  $ U(\cdot)= - \infty $ if $ e^{[k,l]}=0 $.} is a logarithmic function, which achieves  proportional fairness among the users. Moreover,  $e^{[k,l]}$ are the resources allocated from AP $ l $ to user $ k $, $\xi_{k} > 0$ is the scalability to flow $e^{[k,l]}$ ~\cite{9521837}, $ r^{[k,l]} $ is the achievable user rate and  $ \rho_l $ is the capacity constraint of AP $ l $. The first constraint satisfies  that AP $ l $ is not overloaded,  and the second and third constraints ensure that each user receives its requirements of resources located within  a certain range, where $ e^{[k]}_{\max} $ and $ e^{[k]}_{\min} $ are  the maximum and minimum resources required by user $ k $, respectively. This optimization problem can be solved through an exhaustive search method, which involves high complexity. In  this sense, full dual decomposition method via the Lagrangian multiplier can be considered to solve this problem  providing sub-optimal resource allocation with less complexity, more details are provided in \cite{AA19901111,9521837,FRLHANZO}. The Lagrangian function of \eqref{OP2}  is   
\begin{equation}
{\begin{aligned}
\label{eq:lag}
\quad & g\left(e,\rho, \lambda , \eta_{1}, \eta_{2} \right) =  \sum_{l\in L}\,
 \sum_{k\in K} \log \left(1+ \xi_{k} ~ e^{[k,l]} r^{[k,l]}\right)\\
 \quad &+ \underbrace{ \sum_{l\in L}\lambda_{l}\left(\rho_l- \sum_{k\in K} e^{[k,l]}\right)}_{\text{constraint 1 in \eqref{OP2}}}+ \underbrace{\sum_{k\in K} \eta_{k,1} \left( e^{[k]}_{\max}- \sum_{l\in L}  e^{[k,l]}\right)}_{\text{constraint 2 in \eqref{OP2}}}\\\quad & ~~~~~~~~~~~~~~~~~~~~~~~~~+\underbrace{\sum_{k\in K} \eta_{k,2} \left(\sum_{l\in L}  e^{[k,l]}-e^{[k]}_{\min}\right)}_{ \text{constraint 3 in~\eqref{OP2}}},
\end{aligned}}
\end{equation}
where $  \lambda_{l}$, $ \eta_{1}  $ and $ \eta_{2} $  are multipliers associated with the first, second and third constraints in \eqref{OP2}, respectively. It is worth mentioning that solving \eqref{eq:lag} requires an iterative algorithm, which is not practical to use in providing instantaneous  estimations of resources  in real time scenarios. In the following, an ANN model is presented to avoid the complexity of the optimization  problem in \eqref{OP2}, while providing a sub-optimal solution.  
\section{Artificial Neural Network}


\begin{figure}[t]
\begin{center}\hspace*{0cm}
\includegraphics[width=0.8\linewidth]{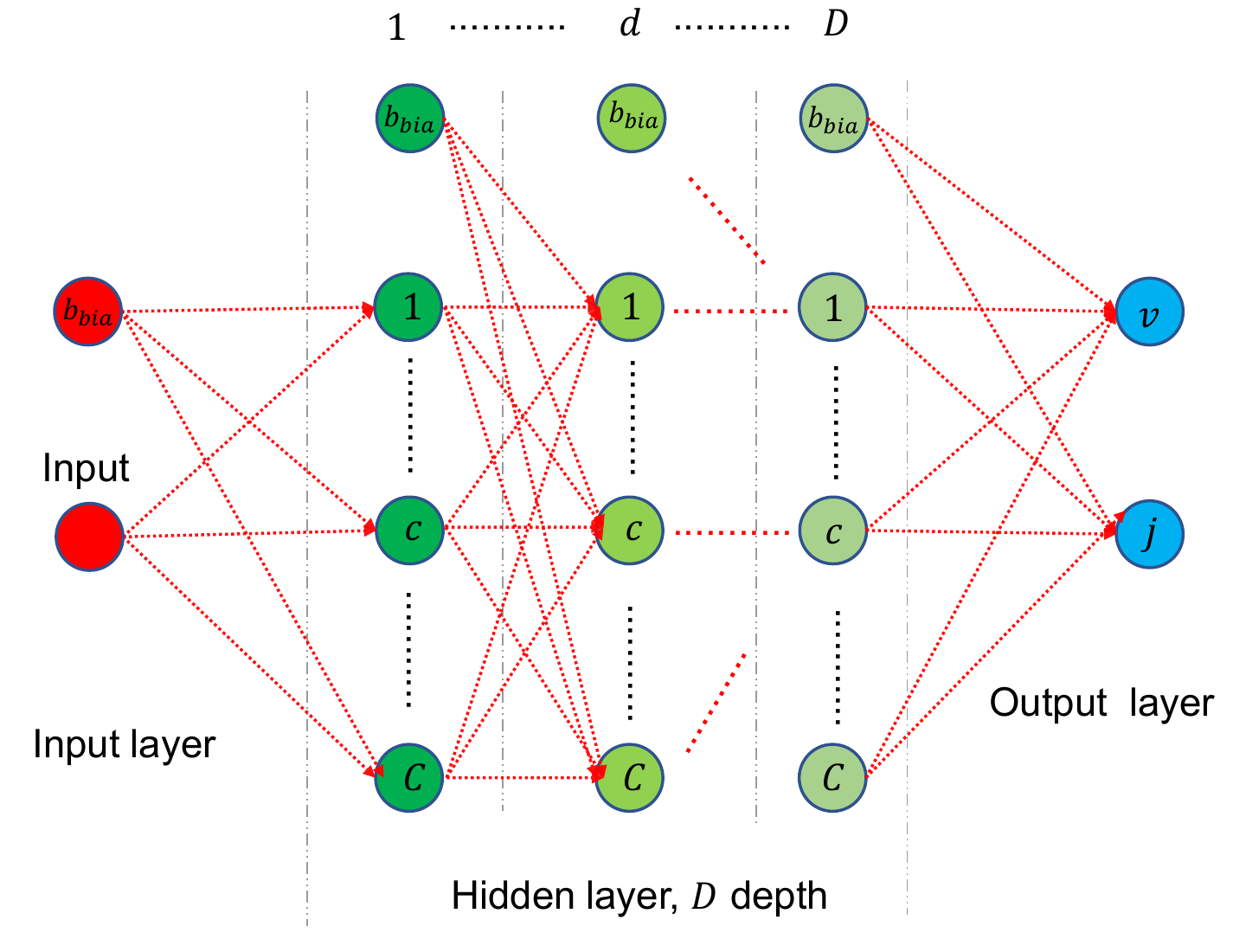}
\end{center}
\vspace{-2mm}
\caption{An ANN model with an input layer, $ D $ hidden layers and two output layers.}\label{nna}
\vspace{-2mm}
\end{figure} 
Generally, an ANN model consists of a series of multiple layers, each  layer is composed of artificial neurons that preform a certain function. An example of the ANN model is presented in Fig.\ref{nna}, for the sake of simplicity. As shown, it has an input layer, multiple hidden layers and an output layer.  The  input layer usually corresponds to information received from the environment  with a bias denoted as $ b_{bia} $, on which mathematical calculations are applied to provide  intended solutions. The hidden layer is composed of  $ D $, $ d=\{ 1, \dots, D\}, $ multiple layers, and each artificial layer $ d $  has a total of $ C $ artificial neurons where each neuron is designed with the ability to perform certain mathematical calculations on its input, so that its output can be fed into the next layer. The output layer of the ANN model consists of two artificial neurons with outputs determined based on information that is fed  from the hidden layer $ D $. It is worth pointing out that in this work, the ANN model must learn the requirements of the users and the  available resources, and then, the output layer gives the resources allocated to each user and the capacity limitation of each AP.   


Focusing on  the $ d $ th layer,   the input and output of this layer are given by 
\begin{equation}
\mathbf{s}_{i,d}= [s_{1,d-1} \dots, s_{C,d-1}]^{T},
\end{equation}
and
\begin{equation}
\mathbf{s}_{o,d} = [s_{1,d} \dots, s_{C,d}]^{T},
\end{equation}
respectively. It is easy to notice that the input of the $ d $ th layer is the exact output $ \mathbf{s}_{o,d-1} $ of layer $ d-1 $. Moreover, the output $ s_{c,d} $, $ c \in C $, is calculated as in the following
\begin{equation}
\label{eq:fuc}
s_{c,d}  =  \kappa_{c,d} [\alpha( \mathbf{s}_{o,d-1},  \mathbf{w}_{c,d}, b_{bia_{c,d}})],
\end{equation}
where  $ \kappa [.] $  is an activation function, $ \alpha (.,.,.) $ is a linear transformation function,  $ \mathbf{w}_{c,d} $ is the the weight vector of  neuron $ n $ in layer $ d$-th, and $ b_{bia_{c,d}} $ is a scalar bias.  Additionally, the linear transformation function $ \alpha (.,.,.) $ of the ANN model  is given by 

\begin{equation}
\alpha( \mathbf{s}_{o,d-1},  \mathbf{w}_{c,d}, b_{bia_{c,d}}) =  \mathbf{w}_{c,d} \circledast \mathbf{s}_{o,d-1}+  b_{bia_{c,d}},
\end{equation}
where $ \circledast $ is the convolution operator  where a convolutional neural network (CNN) is considered in this work due to its relatively high speed and efficiency. According to the process above, the outputs of the ANN model in Fig.3, $ v $ and $ j $, are given by 
\begin{equation}
\label{eq:fuc}
v  =  \kappa_{v,o} [\alpha( \mathbf{s}_{o,D},  \mathbf{w}_{v,o}, b_{bia_{v,o}}) ],
\end{equation}
and
\begin{equation}
\label{eq:fuc}
j  =  \kappa_{j,o} [\alpha( \mathbf{s}_{o,D},  \mathbf{w}_{j,o}, b_{bia_{j,o}}) ],
\end{equation}
respectively.
At this point, we aim to find the mapping $ f(\mathcal{V};.) $ of the ANN model that guarantees allocating the resources  based on the requirements of the users and the available resources,  where $ \mathcal{V}  $ is defined as a set of weight terms working as bridges among the  vital parameters of the optimization problem that maximize the sum rate of the network. The optimalaity of $ \mathcal{V}  $ can be ensured following a training process presented in the next sub-sections.

\subsection{Dataset generation}
A dataset obtained from solving the main problem can be used for training the ANN model to choose 
 the accurate set of parameters, i.e., $  \mathcal{V^{*}}   $,  that  provide a sub-optimal solution in a real time scenario. Our system model is composed of a number of APs serving  multiple users distributed on the receiving plane. Each user sends its resource requirements  through a WiFi link or low data rate diffuse optical wireless link
 to all the available APs. 
 
 In this context, our aim is to get a set of data points  given by $ N $, each $ n $ corresponds to $ K $ users sending their resources requirements  located within the range  $  e^{[k]}_{min}\leq e^{[k]} \leq  e^{[k]}_{max} $ to $ L $ APs. It is worth mentioning that the resources allocated for a certain user  $ e^{[k]} $ is given by $ \sum^{L}_{l=1} e^{[k,l]}$, and its value differs from one user $ k $  to another $ k' $ according to the activity of a user at a given time, i.e., $ e^{[k]} \neq e^{[k']}, k \neq k' $. Moreover, the value of $ e^{[k,l]} $  allocated to user $ k $ from  AP $ l $ is determined in accordance to the  capacity limitation of  that AP, and it is  different from the value of $  e^{[k,l']}  $ allocated  to the same user from another  AP $ l \neq l' $ . 

The resources  required by each user are based on satisfying its requirements as well as maximizing the sum rate of the users. That is, each AP $ l $ solves the following equation independently to maximize its own utility function

\begin{multline}
\max_{e} \{ \sum_{k \in K} \log (1+ e^{[k,l]} r^{[k,l]})- \\ \lambda_{l} \sum_{k \in K} e^{[k,l]}- \sum_{k \in K} (\eta_{k,1}-\eta_{k,2}) e^{[k,l]}\}.
\end{multline}
The multipliers $\lambda  $, $ \eta_{1} $ and $ \eta_{2} $ work  corresponding to the capacity limitation and user-requirements constraints, respectively (see \eqref{OP2} and \eqref{eq:lag}). To solve this problem, the optimal resources allocated to each user by an AP are determined for fixed values of the multipliers  by applying the Karush - Kuhn - Tucker (KKT) conditions \cite{dfdf}. Then, an updating process for the multipliers  is preformed to satisfy the requirements of users by increasing the resources allocated to each user towards the maximum value $ e^{[k]}_{max} $ if there are sufficient resources, otherwise the multipliers are updated to decrease the resources allocated to the minimum value $ e^{[k]}_{min} $ \cite{9521837}, i.e.,  

\begin{equation}
\label{var}
\lambda_{l}(i+1)= \left[\lambda_{l}(i)-\Omega_{\lambda}(i) \left(\rho_l-\sum_{k\in K } e^{*[k,l]}\right) \right]^{+},
\end{equation}
\begin{equation}
\label{lam}
\eta_{k,1} (i+1)=\left[\eta_{k,1}(i)-\Omega_{\eta_{1}}(i)\left( e^{[k]}_{max}-\sum\limits_{l=1}^{L}e^{*[k,l]}\right)\right]^{+},
\end{equation}
\begin{equation}
\label{nu} 
\eta_{k,2} (i+1)=\left[\eta_{k,2}(i)-\Omega_{\eta_{2}}(i)\left(\sum\limits_{l=1}^{L}e^{*[k,l]}- e^{[k]}_{min}\right)\right]^{+},
\end{equation}
respectively, where $ i$th denotes the iteration of the gradient algorithm, $ [.]^{+} $ is a projection on the positive orthant
to account for considering the fact that we have $ \lambda, \eta_{1}, \eta_{2} \geq 0 $. Furthermore, $ \Omega_{j}(i) $ , $ j\in \{ \lambda, \eta_{1}, \eta_{2}\} $, is the step size at a given $ i-$th iteration that is taken in the direction of the negative gradient for the multipliers  $ \lambda $, $ \eta_{1} $ and $ \eta_{2} $. 
After running the optimization problem for different  user activities, the maximized rates are recorded for training the ANN model.

\subsection{ANN Implementation}
The implementation of the ANN model  can avoid the high complexity of solving the optimization problem in \eqref{OP2}. Notice that, generating the dataset for learning purposes involves also high complexity. However, it is  an offline process, and then, the results can be  recorded  for use in satisfying  the requirements of users instantly in real time scenarios. 
\subsubsection{Offline phase}
The ANN model must be trained over the  dataset generated to find the unknown mapping  between  user-requirements on one side  and resource  allocation and the AP capacity limitations on the other side. In other words, the  set of optimal weight terms  $  \mathcal{V^{*}}  $ must be found in order to make the input and output of the ANN model relevant. In particular, the output layer of the ANN model applied to solve our problem in \eqref{OP2}  estimates $ \widehat{\mathbf{e}}= [\widehat{{e}}^{[1]}, \dots, \widehat{{e}}^{[k]}, \dots, \widehat{{e}}^{[K]}] $, where  $\widehat{{e}}^{[k]}=\sum^{L}_{l=1}\widehat{e}^{[k,l]} $ is the estimation of the optimal resource allocated to user $ k $, which is  given by  $ {e^{*[k]}}=\sum^{L}_{l=1} e^{*[k,l]} $. Notice that, the ANN model estimates also the capacity  limitations of the APs due to the fact that the optimal resource allocation is determined under the first constraint in the original optimization problem (see equation \eqref{OP2}).
  
Let us focus on having  a training  dataset that contains $ N $ data points.  At data point $ n $, $ n=[1, \dots, N] $, the optimal resource allocation is given by $ {\mathbf{e^{*}} (n)} $ for  the training  input denoted as $ {\mathbf{v}} (n) $, while the estimation for this  data point is given by $ \widehat{\mathbf{e}}(n)$. In this sense,  we train the ANN model to choose the optimal set of weight terms that minimizes a certain loss function between the optimal and estimated resources, i.e.,  
\begin{equation}
\label{ANN train}
\\ \min_{\mathbf{W}} \frac{1}{N} \sum^{N}_{n=1} \ell (\widehat{\mathbf{e}}(n), \mathbf{e^{*}}(n)).
\end{equation}
where $ \ell(.,.) $ is the mean-square-error (MSE) function. By solving equation \eqref{ANN train}, the ANN is trained to find   sub-optimal resource allocation for several data points even if these are not included in the training dataset. In a real time scenario, the requirements of the users might change from one time to another, and by sending the new requirements to the ANN model at a given time, an instantaneous solution can be provided with low complexity.

\subsubsection{Real time phase}
In the offline phase, the ANN model is trained to find the optimal weight terms. Subsequently, the ANN model is deployed at the APs to perform resource allocation in an online phase where each  user requires a certain amount of resources based on its activity. The requirements of the users  and the available resources are fed into the ANN model in order to determine the resources allocated to each user from  the whole set of the APs. Notice that, the overall resources allocated for user $ k $ must satisfy the condition $ e^{[k]}_{min}\leq e^{[k]} \leq  e^{[k]}_{max} $. Therefore, during the process of calculating the resources of each user,  if one AP $ l $ has sufficient resources compared to the other APs, that AP allocates more resources to user $ k $ in order to increase the resources $ e^{[k]} $ towards the maximum value $ e^{[k]}_{max} $, and therefore, maximizing the sum rate of the users, otherwise  the resources $ e^{[k]} $ decease to the minimum value $ e^{[k]}_{min} $. Finally, the APs and users update their multipliers  according to the outputs of the ANN model (see equations \eqref{var}, \eqref{lam} and \eqref{nu}), so that, if any user is not satisfied with its allocated resources, the AP with low multiplier allocates more resources to that user. 
Moreover, if any user changes its activity, a new set of user-requirements must be fed into  the ANN model for new resource allocation among users.

\begin{table}
\centering
\caption{Simulation Parameters}
\begin{tabular}{|c|c|}
\hline
Parameter	& Value \\\hline
VCSEL Bandwidth	& 5 GHz \\\hline
VCSEL Wavelength  & 830 nm \\\hline
VCSEL beam waist & $ 10-30~ \mu $m \\\hline
Physical area of the photodiode	&15 $\text{mm}^2$ \\\hline
Receiver FOV	& 45 deg \\\hline
Detector responsivity 	& 0.53 A/W \\\hline
Gain of optical filter & 	1.0 \\\hline
Laser noise	& $-155~ dB/H$z \\\hline
ANN model & CNN \\\hline
Number of hidden layers & $ D=3 $ \\\hline
Dataset size & $ N= 10^{4} $, $ N=5000 $\\\hline
Training & $ 90\% $ of $ N $\\\hline
Validation & $ 10\% $ of $ N $\\\hline

\end{tabular}
\end{table}

\begin{figure}[t]
\begin{center}\hspace*{0cm}
\includegraphics[width=0.7\linewidth]{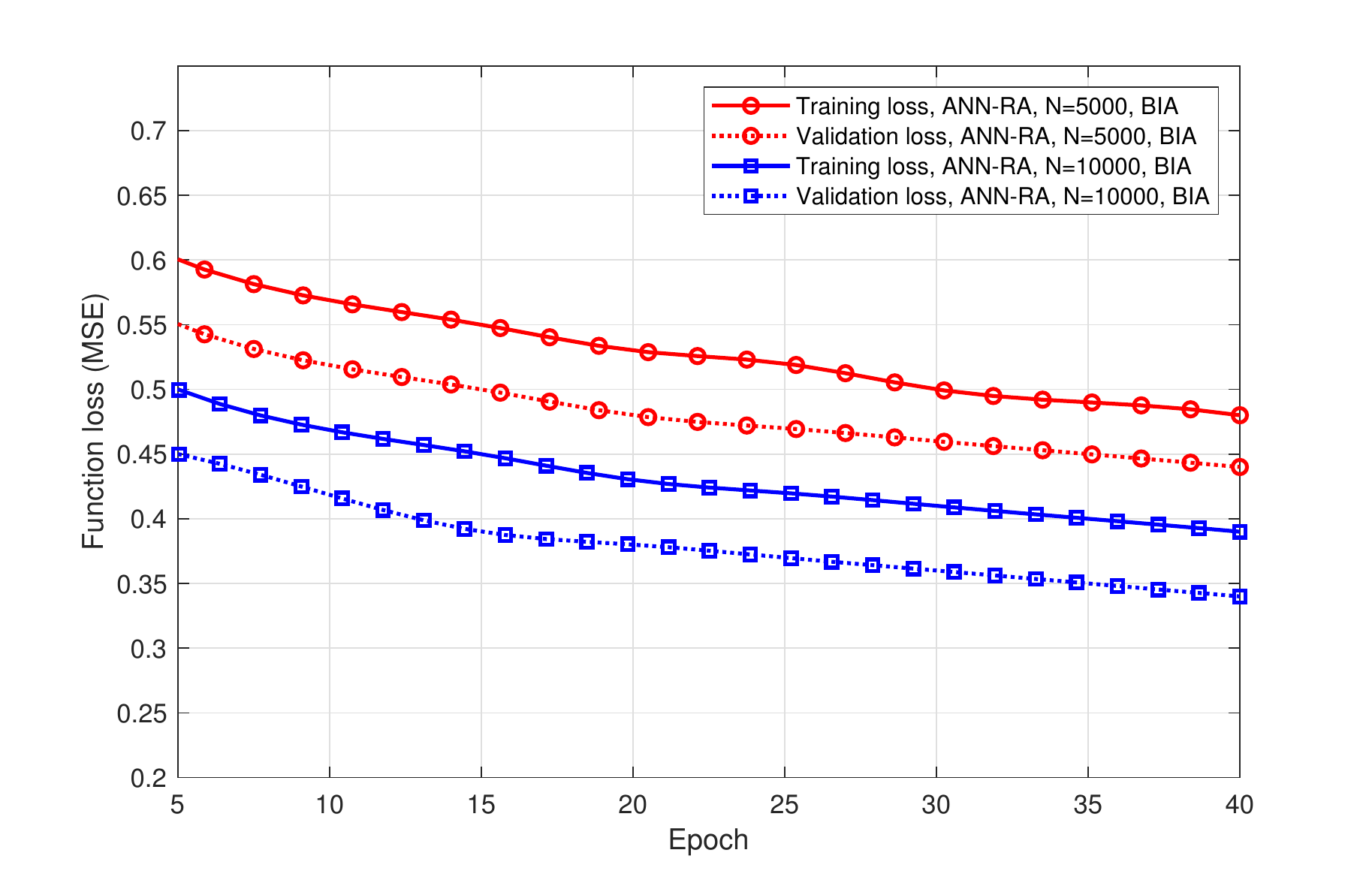}
\end{center}
\vspace{-2mm}
\caption{Training and validation  of the ANN model for two different sizes of dataset.}\label{Ep}
\vspace{-2mm}
\end{figure}

\section{PERFORMANCE EVALUATIONS}
 An indoor environment with  5m$ \times $ 5m$  \times $ 3m dimensions is considered to test the performance of the proposed ANN model. On the ceiling, $ L=16  $ VCSELs are deployed with a uniform distribution to form an array of VCSELs serving $ K=10 $ active users with different requirements at a given time. These users are distributed randomly  on a receiving plane with a 2.15m distance from the ceiling, and each user is equipped with a reconfigurable detector that consists of $ M $ photodiodes, and has the ability to provide $ L $ preset modes in order to apply BIA successfully, mores details are in \cite{AA19901111,8636954}. All the other simulation parameters are listed in Table 1.

The accuracy of  the ANN model is shown in Fig.~\ref{Ep}, in terms of training and validation losses  versus a set of epochs. The MSE of the training loss over a $ N=5000 $ dataset size is 0.6 at epoch 5, and it starts decreasing with  the number of epochs, while the validation loss is 0.55 for the same dataset size at epoch 5. On the other hand, the accuracy  of the ANN model increases with the size of the dataset, where the MSE losses of the training and validation processes over a $ N=10000 $ dataset size  are $ 0.5 $ and $ 0.45 $ at epoch 5, respectively.  Notice that, the ANN model for both dadaset sizes is not overfitting, and  an acceptable solution can be obtained even if the ANN model is fed with information that are not included in the dataset used for the training process.   
Therefore, the ANN model is validated to provide sub-optimal solutions in real time scenarios where the requirements of users and the available resources of the network might change multiple times in a few seconds.
\label{sec:Pcom}

\begin{figure}[t]
\begin{center}\hspace*{0cm}
\includegraphics[width=0.7\linewidth]{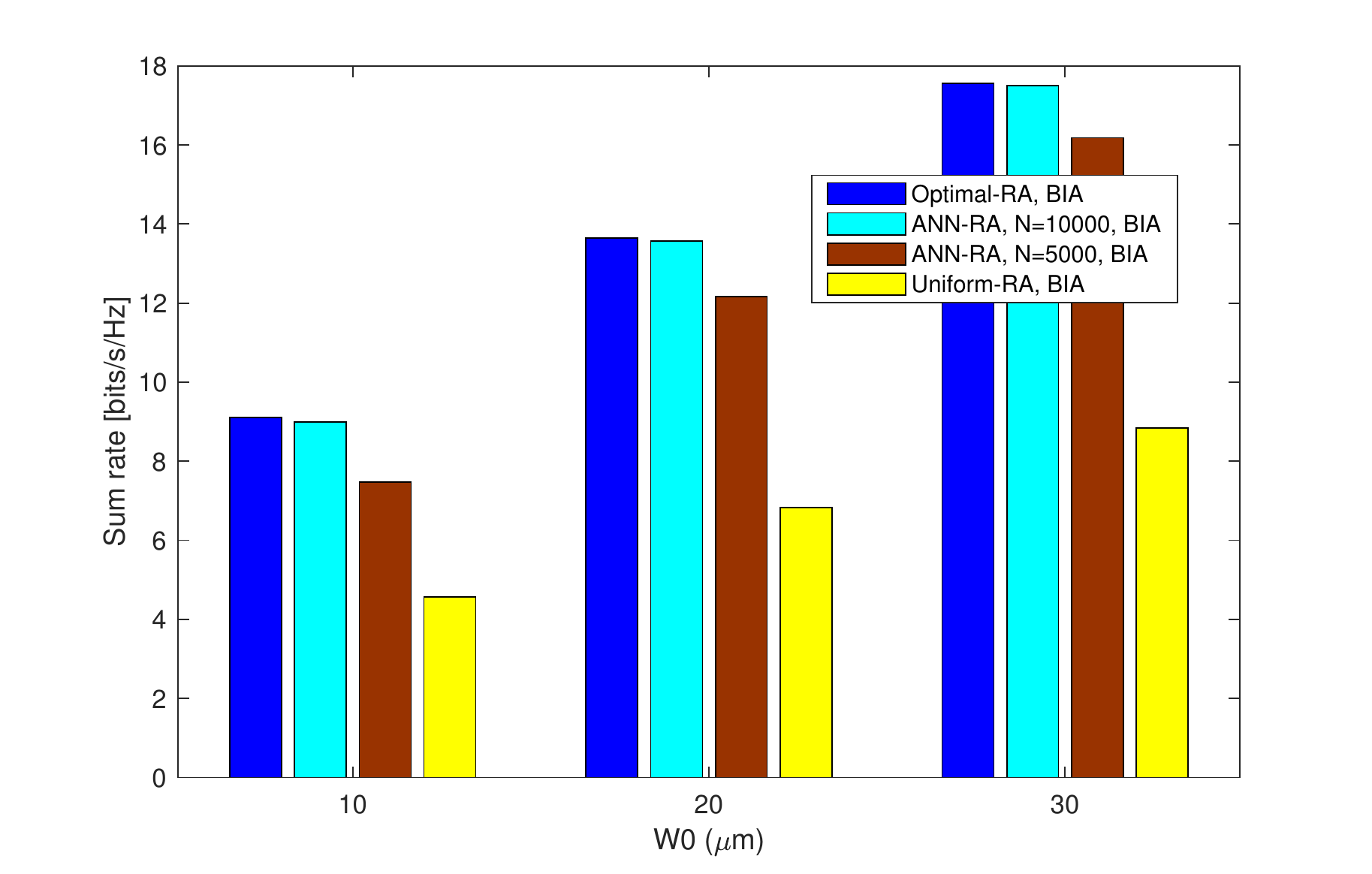}
\end{center}
\vspace{-2mm}
\caption{Achievable sum rates of BIA for the  optimization techniques considered versus different values of the VCSEL beam waist $ W0 $.}\label{rate}
\vspace{-2mm}
\end{figure}

In Fig.~\ref{rate}, the sum rate is depicted  against the beam waist of the VCSEL, considering two different sizes of the dataset used for training the ANN model. It can be seen that the ANN model trained over $ N=10000 $ achieves a solution with high accuracy compared with $ N=5000 $ in all the scenarios considered. Compared with solving the main problem in \eqref{OP2}, the ANN model provides a significant sum rate  close to the optimal solution, and therefore, using the proposed ANN model is beneficial in solving optimization problems of high complexity in  real time. Moreover, allocating the resources based on the proposed utility function results in a higher sum rate than simply dividing the resources of the network among the users regardless of their demand, where some users might waste the resources  allocated due to their low rate applications used at a given time. The figure further shows that increasing the beam waist of the VCSEL enhances the sum rate achieved for all the optimization techniques, which is due  to the fact that the received power increases with increase in the beam waist where the transmitted power is considerably focused towards  the users as the illuminated area of the VCSEL gets more confined.       

\begin{figure}[t]
\begin{center}\hspace*{0cm}
\includegraphics[width=0.7\linewidth]{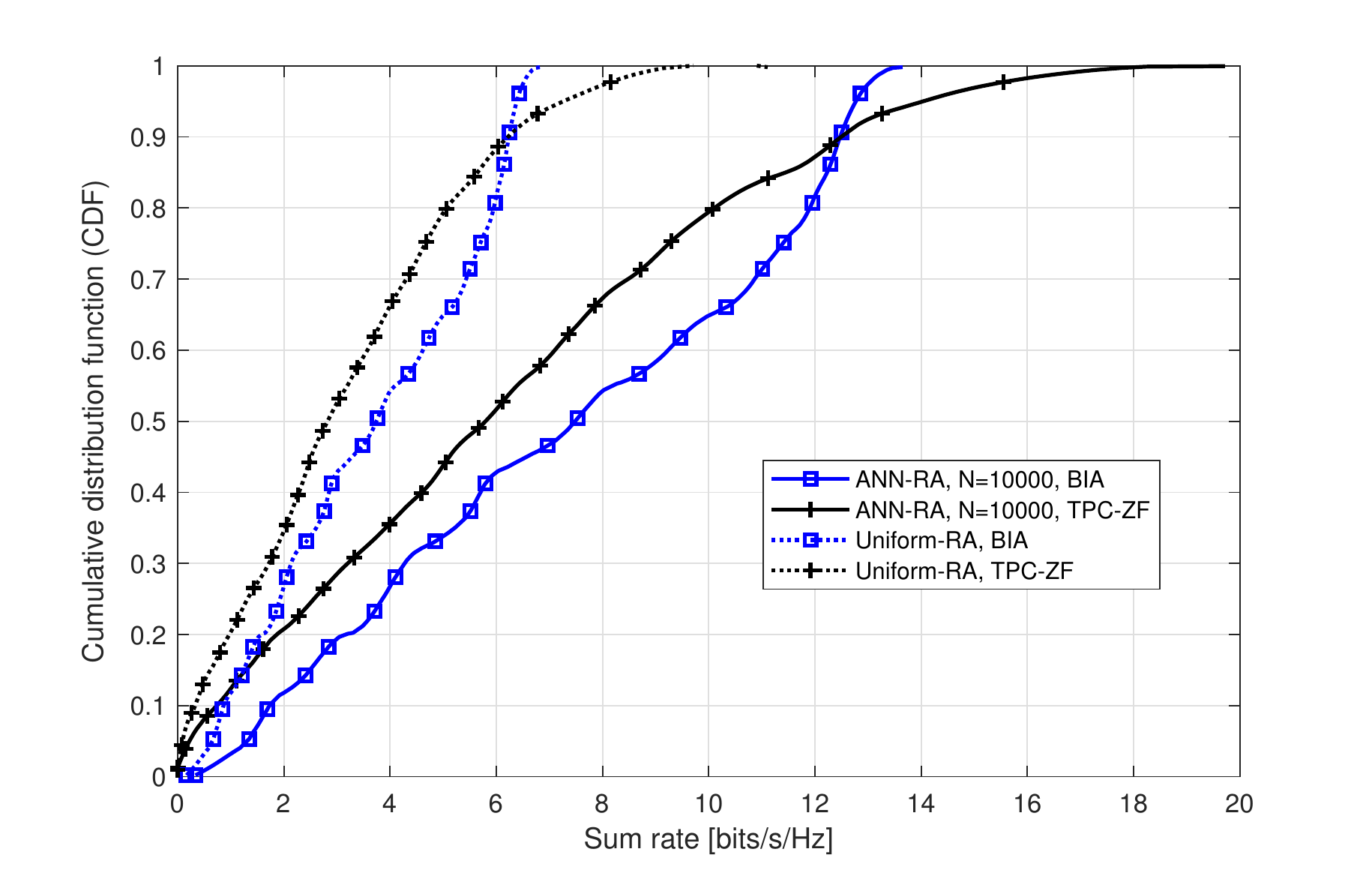}
\end{center}
\vspace{-2mm}
\caption{CDF of the sum rate for the optimization problem solved using the ANN mode compared to uniform scheme.}\label{cdf}
\vspace{-2mm}
\end{figure}

In Fig. \ref{cdf}, the cumulative distribution function (CDF) is shown for the sum rate of BIA compared with ZF. The performance of BIA is superior to ZF in both  resource allocation scenarios using the ANN model and the uniform scheme. It is worth mentioning that BIA satisfies the non-negativity of the transmitted signal naturally due to its positive precoding matrix given by 0 and 1, and therefore, applying a DC bias current, which might cause clipping distortion to the transmitted signal, is avoided. In contrast, the performance of ZF is limited  due to the characteristics of the optical channel where the negative values of the transmitted signal must be avoided strictly by applying a DC bias current \cite{AA19901111,8636954}. Finally, the  ANN model trained over an $ N=10000 $ dataset size achieves higher sum rate than the uniform resource  allocation scheme. 
\section{CONCLUSIONs}
\label{sec:CONCLUSIONs}
 In this paper, an optimization problem is formulated in a laser-based wireless network with an objective function that aims to allocate resources based on the requirements of users.  This problem can be solved by exhaustive search, which involves high complexity, or by distributed algorithms via Lagrangian multipliers, which requires an iterative algorithm that might consume time.  Therefore, an ANN model is introduced to solve the problem in a real time scenario with low computational time. We first generate a dataset from solving the main problem in an offline phase, and then, the ANN model is  trained over this dataset to choose an optimal set of weights that minimizes a certain loss function. After that, the  trained ANN is implemented in a real time scenario to allocate resources among users considering their requirements and the available resources of the network. The results show the optimality of the ANN model where an instantaneous sub-optimal solution close to the optimal one is provided.

\bibliographystyle{IEEEtran}
\bibliography{IEEEabrv,mybib}
\end{document}